\documentclass[twocolumn,tightenlines,showpacs,aps]{revtex4}
\usepackage{graphicx}

\begin{document}

\title{Bi-Analyte Surface Enhanced Raman Scattering for unambiguous
evidence of single molecule detection}

\author{E. C. Le Ru} \email{Eric.LeRu@vuw.ac.nz}
\author{M. Meyer}

\author{P. G. Etchegoin} \email{Pablo.Etchegoin@vuw.ac.nz}

\affiliation{The MacDiarmid Institute for Advanced Materials and Nanotechnology\\
School of Chemical and Physical Sciences\\ Victoria University of Wellington\\
PO Box 600, Wellington, New Zealand}

\date{\today}

\begin{abstract}
A method is proposed to pin down an unambiguous proof for single
molecule surface enhanced Raman spectroscopy (SERS). The
simultaneous use of two analyte molecules enables a clear
confirmation of the single (or few) molecule nature of the
signals. This method eliminates most of the uncertainties
associated with low dye concentrations in previous experiments. It
further shows that single-molecule signals are very common in
SERS, both in liquids and on dry substrates.
\end{abstract}


\maketitle

\section{Introduction}

Surface Enhanced Raman Scattering (SERS) was discovered in the
1970's and was immediately the subject of intense research up to
the 1980's. The possibility of observing Raman signals, which are
normally very weak, with enhancements of the order of $10^6-10^8$
had interesting applications, in particular in analytical
chemistry. However, the lack of reproducibility, of adequate SERS
substrates, and of understanding of the SERS enhancement
mechanisms hampered progress. An in-depth review of these early
studies can be found in Ref.\ \cite{Moskovits85}. Two independent
reports \cite{Nie97, Kneipp97} in 1997 of the observation of
single molecule emission under SERS conditions triggered a renewed
interest in this technique. It meant that enhancement factors
might be as large as $10^{15}$ to compensate for the intrinsically
small Raman cross section, and that SERS probes could potentially
replace fluorescent ones in several applications; for example in
biology. Some advantages of SERS over fluorescence are its higher
spectral specificity and the possibility of using infrared
excitation (important in living tissues for example). Studies of
single-molecule SERS (SM-SERS) could also lead to a better
understanding of the SERS effect itself. However, SM-SERS has
encountered many of the same problems: fluctuations,
non-reproducibility, and a lack of understanding of the origin of
the sites suitable for SM-SERS. Another major problem is the fact
that single-molecule emission has so far been inferred from
indirect evidence, casting doubts over the reality of SM-SERS in
the first place and giving rise to alternative explanations. This
is perfectly summarized in the title of a discussion
\cite{Chimia99} by several of the authors of the first reports of
SM-SERS \cite{Nie97, Kneipp97}: "Single Molecule Raman
Spectroscopy: Fact or Fiction?". They stress that the inference of
SM-SERS from their results is not straightforward, and that
although many evidence support it, it does not constitute a proof
in the absolute sense. In this report, we propose and apply a
technique, Bi-Analyte SERS (BiASERS) which provides a much more
direct evidence for SM-SERS. The principle is to carry out SERS
measurements on a mixture of two different analytes (dyes). This
alternative presents several advantages over previous techniques:
firstly, it does not rely on ultra-low concentrations of analytes
and on the observation of rare events (such as that of a `hot'
particle \cite{Nie97}). Secondly, the simplicity of the experiment
and its interpretation makes it a more direct and convincing
evidence of SM-SERS. Finally, it can be applied to most SERS
substrates (dry or colloidal) to test for the presence of SM-SERS.
Using BiASERS, we show that SM-SERS is, in fact, quite common and
gives information on the nature of the active sites, the so-called
`hot spots'.

\section{Previous work}

In order to put this work in context, we first review the various
evidence for SM-SERS. By far, the largest group of evidence comes
from studies of SERS on dry silver colloidal particles
\cite{Nie97, Michaels99, Xu99, Weiss01}. Silver colloids mixed
with dyes are immobilized after drying or spin-coated on a
suitable substrate. The dye concentration is chosen so that there
are a small number of dyes per colloids. SERS signals from
individual colloids or clusters are then collected and analyzed.
The single molecule nature of these signals is inferred mainly
from two characteristics:
 \begin{itemize}
\item
 Firstly, the low dye concentration suggests that,
statistically speaking, there cannot be much more than one dye per
colloid \cite{Nie97, Michaels99,Xu99}. However, it was
acknowledged early that these concentration estimates do not
necessarily provide a satisfactory proof \cite{Nie97,Chimia99}.
They are indeed prone to large errors: colloid concentration is
usually estimated from a knowledge of Ag mass used during
preparation \cite{Lee82} and an estimate of their average size (or
volume). Any non-reacted Ag, or the presence of a small number of
much-larger-than-average particles could lead to an overestimation
of colloid concentration and therefore an underestimate of
dye:colloid ratio. Moreover, dye concentrations below 1\,nM
require particular care to avoid contamination, wall adsorption,
and dilution errors \cite{Hildebrandt84}. A further source of
uncertainty is the fact that only a small proportion of colloid
aggregates (so-called `hot' particles) seem to give rise to SERS
signals. This means that there is a possibility that those active
aggregates are the ones who have adsorbed a larger than average
number of dyes because, for example, they present a larger surface
area, or are composed of many individual colloids.
 \item
Secondly, these SERS signals exhibit strong fluctuations, both in
intensity and spectral shape, along with blinking (alternating
on/off periods). These are usually considered a characteristic of
SM-emission \cite{Michaels99,Weiss01,Maruyama04}. However, such
fluctuations are often observed in SERS, even in conditions of
high dye concentration, where the signal is not believed to
originate from single molecules. They were also observed in the
SERS spectra of residual amorphous carbon on the colloids, and
attributed to ongoing photo-induced chemical reactions on the
surface, such as photo-oxidation \cite{Kudelski00, LeRuCPL04}.
Blinking can be observed in SM fluorescence as a result of
photo-bleaching. However, fluorescence-induced photo-bleaching
requires exciting the molecule to its first excited state. It is
believed that photo-bleaching is minimized under SERS conditions
because of very fast energy transfer to the metal. Other
photo-induced effects in SERS can be desorption, molecule
dissociation, or modification of the silver configuration itself
(through photo-oxidation for example). These can be induced
directly by light or indirectly by photo-induced thermal heating
of the metal substrate \cite{LeRuFD05}. The difference with
photo-bleaching is that these effects are cooperative, i.e. they
are likely to affect all molecules at the same time (for example
when the metal reaches a critical temperature). Therefore,
blinking in SERS cannot be invoked as an unambiguous proof of
single molecule emission.
\end{itemize}

Other indirect evidence were put forward, for example by studying
the polarization property of the SERS signal \cite{Nie97}, but
they rely on a theoretical understanding of the details of the
SERS mechanism, which is still open to debate. In particular, the
polarization selection rules of a molecule in close proximity to
surface plasmons can be severely modified with respect to the bare
Raman tensor polarization selection rules; an effect which has a
longstanding history in SERS and related techniques
\cite{Moskovits01}.

Another type of SM-SERS study was carried out in liquid colloidal
solution \cite{Kneipp97}. The evidence for SM detection was based
mainly on the ultra-low dye concentration (33\,fM). It was further
supported by the observation of a Poissonian distribution of the
SERS intensities. However, the main problem with this type of
experiments is that the small number of events (100) is not
significant enough to rule out other distributions. For example, a
sample of 100 intensities following a log-normal distribution
often exhibits oscillations similar to that shown in Ref.
\cite{Kneipp97}. Moreover, as pointed out in Ref. \cite{Kneipp97},
such a Poissonian distribution would require a very large
uniformity in the SERS signals (or enhancements), which is not
common in SERS experiments, and that nearly every single molecule
in the scattering volume would be detected. As pointed out
recently \cite{Moskovits02}, this is in contradiction with the
findings of the other original report of SM-SERS where only a
small number of `hot' particles gave rise to SM-SERS \cite{Nie97}.
The convolution problem of the enhancement factor and the dye
concentration is very difficult to break in these approaches.

Despite the uncertainties, this body of evidence has led to the
general acceptance that SM-SERS is a real phenomenon, in
particular on dry substrates, even if no absolute proof was
available. The consensus is that there are very few active
particles ($\approx 0.1\%$) \cite{Nie97} capable of producing SM
signals. In addition, there are on these particles only a few site
capable of providing sufficiently large enhancement for SM-SERS.
These sites are typically at the junction between two closely
spaced colloids \cite{Xu99, Michaels00}, and corresponds to only
0.01 -- 0.1\% of the surface area of a dimer. Such an
interpretation is supported by theoretical studies
\cite{Corni02,Xu03}. One therefore easily sees the conundrum of
SM-SERS: how can we place a molecule at these rare active sites
and be sure that the signal comes from this molecule only? All
approaches so far have relied on low dye concentrations, which
more or less ensures that only a few dyes are observed at a time.
However, the probability of having a molecule at an active site
then becomes very small, leading to unreliable statistics, and
making the SM interpretation of the signals very difficult.

Our approach is exactly the opposite. We use a relatively high
concentration of dyes to ensure that there is indeed at least one
molecule at most active sites. In order to confirm the SM nature
of the signal, we simply use a mixture of two dye molecules
(BiASERS). Because of the large number of molecules, the SERS
signal should in principle always be a mixture of these two dyes.
The observation of a SERS signal of purely one type of dye (say
Dye 1) is a clear evidence that it comes from a very small number
of molecules. For example, if the signal originates from exactly 5
molecules, the probability of it being purely Dye 1 is only 1/32,
going down to 1/1024 for 10 molecules. The main advantage of this
method, in addition to its simplicity, is that it enables to probe
every active sites and it does not rely on an accurate estimate of
the dye:colloid ratio.

\begin{figure}
\centering{
 \includegraphics[width=8cm]{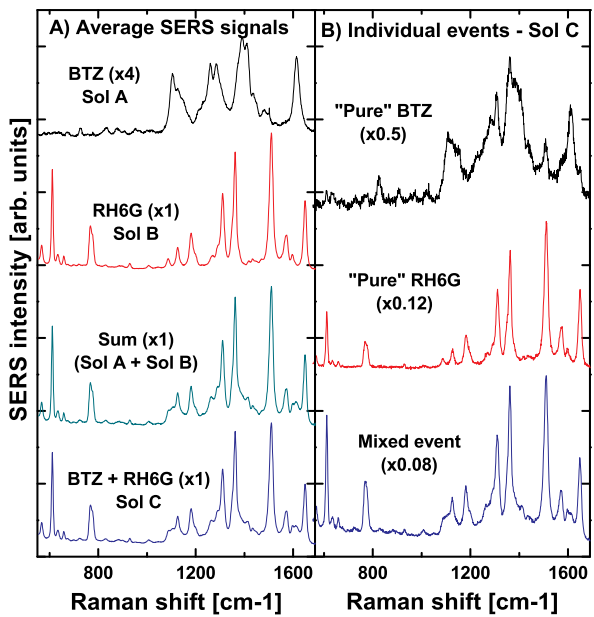}
} \caption{(A) Average SERS spectra from solutions A (100nM BTZ),
B (100\,nM RH6G), and C (100\,nM of each dye). Also shown is the
sum of spectra from A and B, which is identical to the spectrum of
solution C within experimental errors. (B) Representative
individual spectra (integration time 0.2\,s) of solution C showing
a `pure' BTZ event ($p_B=0.91$), a `pure' RH6G event ($p_B=0.07$),
and a mixed event ($p_B=0.5$). Arbitrary scale is the same on both
sides. The `pure' BTZ event still show very small peaks from RH6G
due to its larger cross section.}
 \label{FigAve100nM}
\end{figure}

\section{Experimental Details}

Our citrate-reduced silver colloid solutions were prepared using
standard procedures \cite{Lee82}. Rhodamine 6G (RH6G) was used as
purchased, while the benzotriazole dye (BTZ) was synthesized
following the procedure described in Ref. \cite{Graham98} (dye \#
2 of this reference). RH6G has been widely used in the past for
SERS and SM-SERS. BTZ was designed specifically for SERS studies.
It is believed to adsorb strongly (through covalent bonding) to
silver \cite{Graham98}. Under standard conditions, the two dyes
show strong SERS spectra, which are easily distinguishable, as
shown in Fig. \ref{FigAve100nM}. Of particular interest are
non-overlapping peaks that allow unequivocal identification of the
different dye species. Raman spectra were acquired using a Jobin
Yvon Labram confocal Raman spectrometer, with a $\times 100$
immersion objective index-matched to water or a $\times 100$
objective matched to air for dry substrates. Excitation was
carried out with a 633\,nm HeNe laser (2\,mW at the sample for
liquids, 0.2\,mW for the Raman map on dry substrates). Solution K
was prepared by mixing 0.5\,mL of colloidal solution with 0.5\,mL
of a 20\,mM KCl solution. The main effect of KCl is to reduce the
Coulombic repulsion between the colloidal particles. Using a
35\,mM KCl solution triggers aggregation and collapse of the
colloids. SERS signals are usually very small at low KCl
concentrations ($<5$\,mM), because they originate from single,
non-interacting colloids. Using 20\,mM KCl (10\,mM final
concentration in solution), the colloidal solution remains stable
for several weeks. However, the Coulombic repulsion is
sufficiently reduced to allow colloids to form aggregates. The
stability of these samples indicate that these aggregates must be
small, most likely pairs of colloids. Such interactions are
necessary to observe large SERS signals. All colloidal solutions
were then prepared by adding a small volume of a dye solution
(BTZ, RH6G, or a mixture of both) to solution K. We used final dye
concentration in the range 20 to 100\,nM. All colloid solutions
were left to rest for several hours or more before measurements.

For experiments on dry colloids, a 40\,$\mu$L drop of such a
colloid+dye solution was deposited and left for one minute on a
positively-charged Polylysine-coated glass slide. Because of the
negative charge on the colloids, clusters close to the Polylysine
surface are immobilized through electrostatic interactions
\cite{Nie97}. The slide was then rinsed with distilled water,
leaving a small density of immobilized clusters for SERS
measurements.

\section{Experimental results}

We first focus on a colloidal solution prepared with a mixture of
equal concentration (100\,nM for each dye) of RH6G and BTZ
(solution C). Control samples with 100\,nM of BTZ only (solution
A) or RH6G only (solution B) were also measured for comparison. We
estimate that this corresponds to at least 1200 dyes of each type
per colloids, much more than in any previous SM-SERS studies. We
also estimate a surface density of around 0.1 dye of each type per
nm$^2$, which indicates that there should not be any steric
hinderance for adsorption (typical surface area of an adsorbed dye
is 1 nm$^2$). There is also on average between 1 and 4 colloids
(or between 0.5 and 2 pairs) at any given time in the scattering
volume. A series of 1000 SERS spectra with 0.2\,s integration time
were collected from each solution and analyzed. The results are
summarized in Fig. \ref{FigAve100nM}. The average spectra of
solution A and B show that RH6G and BTZ have clearly
distinguishable SERS spectra. The RH6G spectrum is stronger due to
its higher SERS cross-section under these conditions. Solution C
shows a superposition of these two spectra, which is identical to
the sum of the spectra from sol A and B, within the experimental
errors. This strongly indicates that the two dyes do not interact
with each other and adsorb on the colloids independently of each
other, as expected from the low surface densities. We now focus on
the analysis of the 1000 individual spectra obtained from solution
C. Most spectra exhibit a good signal, indicating that interacting
colloids are on average always present in the scattering volume.
We observe fluctuations in intensity and spectral shape. These are
attributed to constantly changing colloid configurations in the
scattering volume because of the unavoidable Brownian motion, and
are not necessarily a sign of SM-SERS. More interesting are the
large fluctuations observed in the relative proportion of signal
from each of the dyes. For example, in Fig. \ref{FigAve100nM}\,(B)
are shown two representative scans where the signals is composed
of purely one or the other type of dye. This, we believe, shows
unambiguously that the SERS signal is dominated by a very small
number of molecules, and represents the simplest and most direct
evidence for SM-SERS. We believe that this is a reliable
experimental proof of an assumption which, although widely
accepted in the literature as possible, is very difficult to
demonstrate experimentally.

\begin{figure}
\centering{
 \includegraphics[width=8cm]{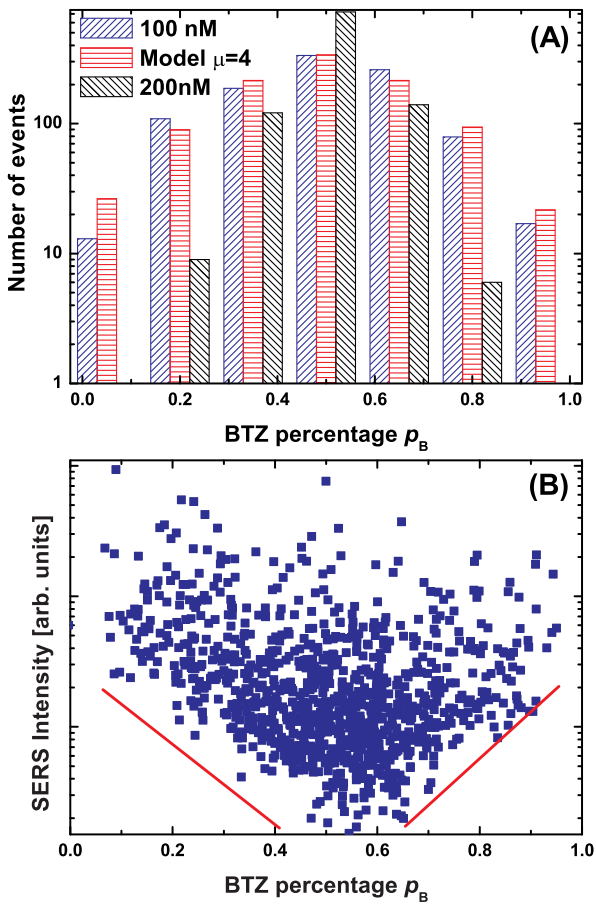}
} \caption{(A) Histograms of the distribution of $p_B$ for
solution C (100\,nM of each dye), for the simple Poissonian model
with $\mu=4$, and for an identical solution with 200\,nM of each
dye.
 A log scale is used to emphasize extreme events.
 (B) Scatter plot of
total SERS intensity versus $p_B$ obtained from the fits. Note
that spectra dominated by one type of dye ($p_B<0.2$ or $p_B>0.8$)
are only observed for high intensity events.}
 \label{FigScatt100nM}
\end{figure}

In addition, the average BiASERS spectrum and the 1000 spectra of
solution C were fitted as a weighted superposition of the average
spectra of BTZ and RH6G (obtained from solution A and B). The fit
for the average spectrum leads to a 1:1 superposition of the
average RH6G-only and BTZ-only spectra, and we assume this
corresponds to a 1:1 dye ratio. The weighted fits therefore enable
to extract a single percentage $p_B$ characterizing the proportion
of the total signal in each spectrum originating from BTZ. If the
enhancement mechanism was uniform, this percentage would
correspond to the proportion of molecules producing the observed
BTZ signal. For example, a fit of the average spectrum gives
$p_B=0.5$ (1:1 dye ratio) even if the integrated intensity is
dominated by RH6G peaks because of its higher cross section. This
procedure, accordingly, acts as a normalization condition for the
different cross sections of the dyes. The statistics of $p_B$ is
illustrated in Fig. \ref{FigScatt100nM} in two forms: (A)
Histograms of the probability distribution of $p_B$, and (B)
correlation plot of $p_B$ with SERS intensity. $p_B$ represents
the proportion of BTZ molecules if the signals from each molecule
was perfectly uniform and equal to the average signal. Because
there are in excess of 1000 molecules of each type on each
colloid, one always expects that $p_B \approx 0.5$, with
negligible fluctuations around this value. Fig.
\ref{FigAve100nM}\,(B) and Fig. \ref{FigScatt100nM}\,(A) show that
this is clearly not the case, with several events where $p_B
\approx 0$ or $p_B\approx 1$. The most likely explanation is that,
at least for these events, the signal is dominated by a few
molecules, those situated at the position of highest enhancements.
 A simple description is to assume that the observed signals
originate only from a given fixed (small) area of large
enhancements , a `hot-spot'. The number of molecules of each type
in this area then follows a Poissonian distribution with the same
average $\mu$ (because dyes are in 1:1 ratio). One can then easily
derive the probability distribution of $p_B$ for a given $\mu$. As
shown in Fig. \ref{FigScatt100nM}\,(A), a value of $\mu\approx 4$
fits well to our experimental results. The statistics of $p_B$
therefore suggests that {\it most clusters exhibit a SERS
enhancement that is sufficient for single molecule detection}
within our integration time of 0.2\,s. Note that we are not
strictly speaking observing SM-SERS here since the signal
originates on average from around 4 molecules, but a signal only 4
times smaller is still well within the detection limit. Reducing
the dye concentration would only result in many events where no
dye are present in the hot-spot, making the statistical analysis
more difficult, like in previous studies. Finally, for larger
values of $\mu$, the distribution of $p_B$ should be increasingly
peaked around 0.5, and the probability of extreme events
($p_B<0.2$ or $p_B>0.8$) should decrease drastically. We clearly
observe this effect experimentally as shown in Fig.
\ref{FigScatt100nM}\,(A) in the histogram for a solution identical
to C, but with doubled concentration for each dye (200\,nM). It is
clear that the occurrence of extreme events has virtually
disappeared. This simple model clearly accounts for the
few-molecule nature of the signals. Moreover, assuming that the
signals originate from a pair of colloids and comparing the value
of $\mu$ to the average number of molecules on this pair, we can
estimate the hot-spot area to be only $\approx 4/2400\approx
0.17\%$ of the total surface area. Such an estimate is much easier
with BiASERS than with conventional low dye concentration methods.
However, this model assumes that the characteristics of the
hot-spots are the same for each event, and that every molecule in
the hot-spot contributes equally to the signal, which are clearly
rough approximations. For example, Fig. \ref{FigScatt100nM}\,(B)
presents a clear indication that the nature of the hot-spots
changes from one event to the other. If large SERS events
corresponded simply to instances where more molecules are present
in the hot-spot area, then these events should be more likely to
be of a `mixed' type, while low intensity events should exhibit
more of the extreme `pure dye' type. Our results suggest the
opposite: `pure' events only occur for high intensity events. {\it
This suggests a strong correlation between the size of a hot-spot
and its enhancement}. Such a correlation is actually predicted by
theoretical studies of the SERS electromagnetic enhancement, where
high enhancement are correlated with strong localization. This is
another characteristic of the effect that is very difficult to
prove experimentally under normal situations.

\begin{figure}
\centering{
 \includegraphics[width=8cm]{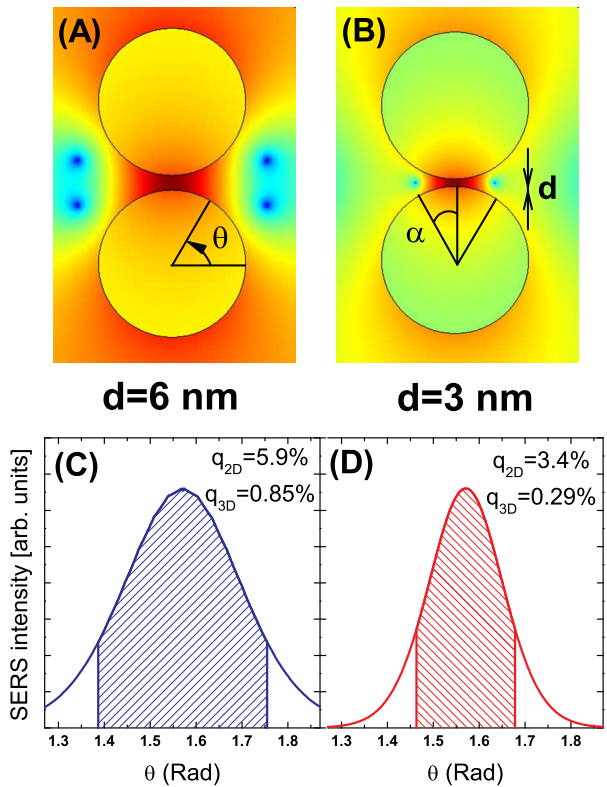}
} \caption{Influence of colloid separation on Hot-Spot
localization. Top: 2D color maps of SERS intensity around the
colloids for separations $d$ of 6\,nm (A) or 3\,nm (B).
Log-scales, different for each plot, are used and red (blue)
indicates regions of high (low) intensities. Bottom: Normalized
SERS intensity on the surface of the bottom disk plotted against
$\theta$, for $d=6$ (C) and 3\,nm (D). Intensity is maximum for
$\theta=\pi/2$, but is more localized for smaller separations. The
greyed area represents $80\%$ of the total area. The Hot-Spot
percentage coverage $q$ is then estimated for the 2D and 3D case.
See text for further details.}
 \label{FigHS}
\end{figure}

It is interesting to relate these results to the standard example
of a hot-spot formed by a pair of closely-spaced colloids. In this
case, the main parameter is the separation $d$ between the
colloids. As $d$ decreases, the SERS enhancement increases and
becomes more localized in between the two colloids. This is
illustrated in Fig. \ref{FigHS} using a simple electromagnetic
model of colloid interaction. The plots were obtained using a 2D
electrostatics approximation and represent the SERS enhancement
factors, approximated by the fourth power of the field amplitude
enhancement, $|E|^4$. The two cylinders have a diameter of 50\,nm
and are separated by $d=6$ or $d=3$\,nm. The optical properties of
Ag were used for the colloids, and the environment is water
($\epsilon_r=1.77$). Excitation is polarized along the dimer axis
at a wavelength $\lambda=500$\,nm, i.e. red-shifted compared to
the single and coupled plasmon resonances of the system. This
simple model is obviously over-simplified, but it clearly
illustrates the point we want to make: the region of highest
enhancement (hot-spot) is increasingly localized for the smallest
separations. In Figs. \ref{FigHS}\,(b) and (c) we show the
normalized SERS enhancements on the surface of one colloid as a
function of angular position, $\theta$ ($\theta=\pi/2$ corresponds
to the dimer axis). For each plot, the colored area corresponds to
$80\%$ of the total area and is a good representation of the
hot-spot (defined here as the region from which $80\%$ of the
total SERS intensity originates). This corresponds to an angle
interval $[\pi/2 - \alpha ; \pi/2+\alpha]$ from which we derive a
percentage $q_{\text{2D}}=\alpha/\pi$. $q_{\text{2D}}$ is the
proportion of the cylinder circumference covered by the hot-spot.
To extend this result to 3D, we assume that the hot-spot
corresponds to a solid angle $\Omega$ defined by a semi-angle
$\alpha$: $\Omega=2\pi(1-\cos(\alpha))$. The hot-spot therefore
covers a proportion $q_{\text{3D}}=\Omega/(4\pi)\approx \alpha^2
/4$ of the total surface area. Despite the simplicity of this
model, the value of $q_{3D}\approx 0.29\%$ obtained for $d=3$\,nm
is of the same order as that inferred from the histogram of Fig.
\ref{FigScatt100nM}\,(B). It is also apparent that the hot-spot
area is correlated with the colloid separation $d$ and therefore
with the total SERS enhancements. This is in agreement with the
correlation observed in the scatter plot of Fig.
\ref{FigScatt100nM}\,(B).

\begin{figure}
\centering{
 \includegraphics[width=8cm]{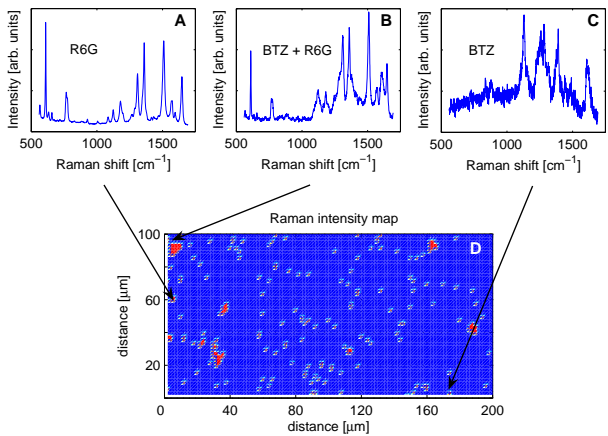}
} \caption{BiASERS for dry colloids. Top: Individual spectra
composed of purely RH6G (A), both RH6G and BTZ (B), and purely BTZ
(C). Note that there is on average in excess of 500 molecules of
each type per clusters. Bottom: $200 \times 100$\,$\mu$m$^2$ SERS
intensity map showing isolated clusters. Red (blue) represents
high (low) SERS intensity regions.}
 \label{FigDry}
\end{figure}

Finally, in order to link this study to previous works, mostly
carried out on dry colloidal particles, we show in Fig.
\ref{FigDry} the results of such an experiment using BiASERS.
There are on average 240 dyes of each type per colloid (dye
concentration is 20\,nM before drying), and therefore more than
500 molecules of each type per cluster, depending on their size.
The Raman map in Fig. \ref{FigDry}\,(D) suggests that all clusters
are active at these concentrations. Moreover, as illustrated in
Fig. \ref{FigDry}\,(A-C), we can again observe signals composed
purely of the SERS spectrum of one type of dye, despite the large
number of molecules of both types. This is again a clear
indication that SM emission is the norm, rather than the
exception. Such a conclusion could only be inferred statistically
in previous low-concentration studies because of the small
probability of finding a molecule at a hot-spot. However, we
insist that the rarity of these events was not due to the small
number of hot-spots or `hot-particles', but to the small
probability of having a molecule there. In other words, most
particles are `hot' provided that a probe molecule is present at
the right position. This is easily achievable in BiASERS (due to
the large concentration of dyes) whilst keeping the ability to
distinguish few-molecules from many-molecules signals.

\section{Conclusion}

In closing, we have proposed and applied a new method to evidence
SM-SERS, using a combination of two SERS probes. Using such a
mixture of two distinguishable probes circumvents many problems
associated with low-concentration studies. It enables to study
most SM-SERS events, instead of a very small number with
unreliable statistics. This technique is simple, unambiguous, and
of wide applicability to various SERS substrates. Our results
readily demonstrate that single-molecule SERS is common, even in
stable colloidal solutions. It could further be used to study a
number of outstanding issues in SERS, which we have only briefly
outlined here. For example, it could shed new light into the
nature of SERS hot-spots themselves, and can also be applied to
determine the SERS cross-sections and enhancements with more
accuracy . All this is necessary for a better understanding of
SM-SERS. This will not, however, solve another major challenge of
SM-SERS in the reverse problem, namely : how we can force a single
available molecule to go to the right position in order to observe
its SERS signal? This will be necessary for some of the most
exciting proposed applications of SM-SERS, such as
single-DNA-molecule sequencing.

\end{document}